\documentstyle[
   aps,
   prl,          
   tighten,      
   epsfig
]{revtex}

\epsfclipon

\begin{document}

\draft

\title{
Novel Mechanism for Single Bubble Sonoluminescence 
}
\author{Boris P. Lavrov}
\address{Faculty of Physics, St.-Petersburg State University, 198904, Russia}
\date{\today}
\maketitle

\begin{abstract}
Careful re-examination of typical experimental data made it possible to show 
that the UV continua observed in multi-bubble (MBSL) and single-bubble (SBSL) 
sonoluminescence spectra have the same physical nature - radiative 
dissociation of electronically excited H$_{2}^*$($a^{3}\Sigma_{g}^{+}$) 
molecules [and probably hydrides of heavy rare gases like 
ArH$^*$($A^{2}\Sigma$)] due to spontaneous transitions between bound and 
repulsive electronic states. The proposed mechanism is able to explain all 
available spectroscopic observations without any exotic hypothesis but in 
terms usual for plasma spectroscopy. 
\end{abstract}


\narrowtext

Sonoluminescence (SL) is one of most exciting features of acoustic cavitation 
- the formation and non-linear oscillation of gas bubbles generated in liquids 
by ultrasound \cite{Setal99}. Bubble collapses lead to enormous local 
pressures (about 1000~atm and more) and temperatures, erosion of hard 
materials, chemical reactions and light emission (SL). Discovered in early 
1930s the SL observed in cavitation clouds is called multi-bubble 
sonoluminescence. MBSL spectra consist of a continuum of unknown origin, 
atomic lines and molecular bands of species connected with host liquids 
\cite{Setal99}. Atomic and molecular emissions give effective temperatures 
$\approx$3000--5000~K \cite{McNDS99}.

\begin{figure}
 \epsfxsize=8.2cm \epsfbox{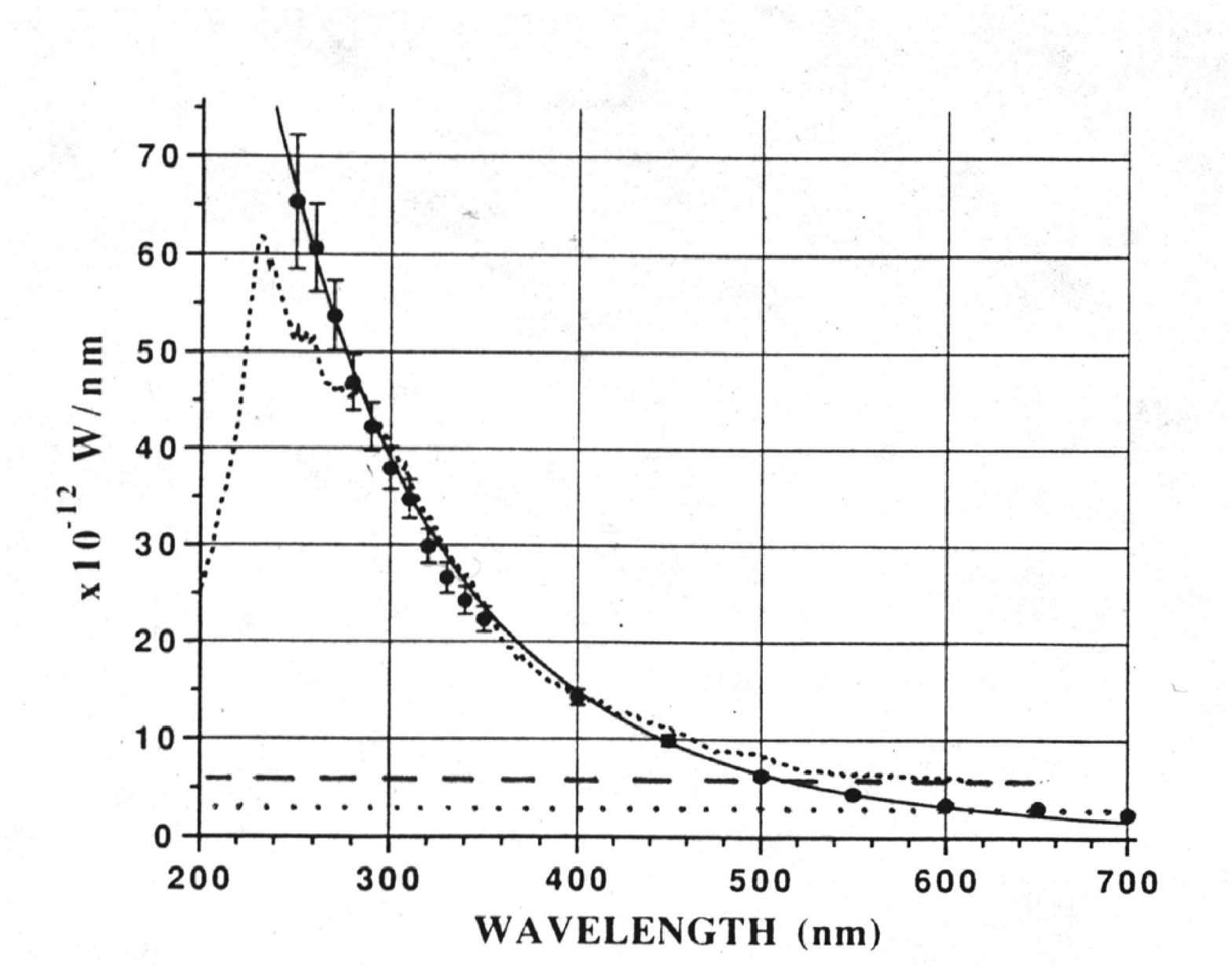}
 \caption{Spectral density of SBSL measured in \protect\cite{HPB92} with two 
different light sources used for the absolute intensity calibration: the 
Deuterium lamp (dotted line) and the quartz tungsten halogen (QTH) lamp 
(points with error bars). The solid line represents blackbody spectrum for 
T=25000 K. The horizontal lines are background levels used in present work.}
 \label{fig:1}
\end{figure}

Suddenly the situation in this traditional branch of non-linear acoustics has 
been changed dramatically after the discovery \cite{GC90} of single bubble 
sonoluminescence - light emission of a single, stable, oscillating bubble 
trapped by acoustic levitation inside an ultrasound resonator \cite{GC90}. 
Such eccentric change in the performance of SL experiment provided unique 
opportunity to study the dynamics of oscillating bubble, obviously masked in 
MBSL. It was shown \cite{P95} that every cycle consists of: 1) relatively slow 
growth of bubble radius up to $\approx$50~$\mu$m, 2) extremely fast implosive 
collapse to $\approx$0.5~$\mu$m, 3) damping and waiting for the expansion 
phase of sound wave in so-called ``dead mode'' with constant radius 
$\approx$5~$\mu$m, corresponding to atmospheric pressure. SL appears as light 
flashes at the moment of the collapse. Further studies (see bibl.~in 
\cite{Setal99,HPB92,P95,C94,BHLPW97}) have shown: 1) water being friendliest 
liquid for SBSL; 2) clockwise regularity of the flashes with $\approx$50~ps 
stability; 3) abnormally short pulse duration ($\approx$50--350~ps); 4) 
featureless spectrum with the intensity increasing to UV (see 
Fig.~\ref{fig:1}), which was fitted as a ``tail of a blackbody spectrum'' with 
abnormally high temperatures $T=25000-100000$~K \cite{HPB92,P95,C94}; 5) 
absence of characteristic emissions associated with host liquid; 6) 
stimulating influence of Ar. Those peculiarities are often presented and 
discussed as mysteries and unknowns of SBSL (see i.e.~\cite{P95,BHLPW97}) not 
only in respected scientific journals, but also in popular ones and newspapers 
as well.

Spectra of MBSL and SBSL measured ``under similar experimental conditions'' 
\cite{MRMMcNS95} are shown in Fig.~\ref{fig:2} (actually the conditions are 
sufficiently different, see below). Such observations are often interpreted as 
evidence that ``these two phenomena are fundamentally different'' 
(``sonochemistry'' and ``sonophysics'') \cite{C94}. Unbelievably high 
temperatures ``observed'' in SBSL experiments were so exciting that the 
mechanism of highly spherical imploding shock wave in SBSL was even considered 
as ``the key to reaching temperatures and densities sufficient to realize the 
fusion of these hydrogen nuclei to yield helium and neutrons'' \cite{P95}. The 
stir stimulated tremendous growth of speculations about the nature of SBSL. 
Many mechanisms have been proposed so far from traditional ones 
(quasi-adiabatic heating, chemiluminescence, electric breakdown, shock wave 
and others) up to such exotic as Schwinger's Dynamical Casimir Effect. 
Although most popular hypothesis is the Bremsstrahlung radiation of electrons 
in dense plasma (see i.e. \cite{Setal99,BHLPW97,Y99}), the real nature of SBSL 
is still an open question \cite{Setal99}.

The main goal of the present work is to propose and consider electronically 
excited H$_{2}^*$($a^{3}\Sigma_{g}^{+}$) molecules [and possibly hydrides of 
rare gases like ArH$^*$($A^{2}\Sigma$)] as light emitters responsible for 
continua observed in both MBSL and SBSL experiments with hydrogen-containing 
liquids. This provides new sight on the well known ``mysteries and unknowns'' 
of the SBSL phenomenon.

\begin{figure}
 \epsfxsize=8.2cm \epsfbox{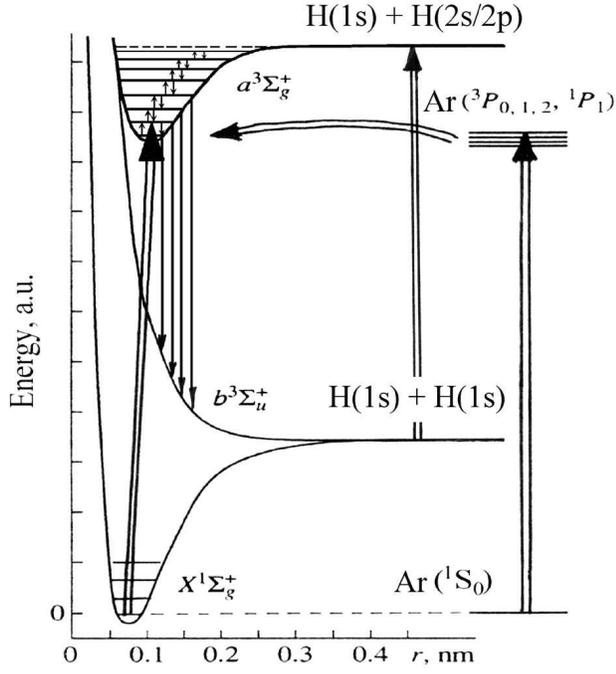}
 \caption{Grotrian diagram of the ground and lowest triplet excited electronic 
states of H$_{2}$ molecule and Ar atom. Arrows indicate $a^{3}\Sigma_{g}^{+},v 
\to$ $b^{3}\Sigma_{u}^{+}$ spontaneous emission transitions. Double arrows 
show some processes of populating of the $a^{3}\Sigma_{g}^{+},v$ vibronic 
states: direct excitation of atoms and molecule and Ar$^* \to$ H$_{2}$ 
excitation transfer.}
 \label{fig:3}
\end{figure}

Perhaps the most simple and natural explanation of the continuum radiation 
observed in SL spectra (never even considered previously) is spontaneous 
emission of hydrogen dissociation continuum appearing due to the 
$a^{3}\Sigma_{g}^{+},v,J \to b^{3}\Sigma_{u}^{+}$ transitions ($v,J$ - 
vibrational and rotational quantum numbers) between lowest triplet bound 
(upper) and repulsive (lower) states (Fig.~\ref{fig:3}). It is a well-known 
spectral feature of hydrogen-containing plasmas widely used in UV spectroscopy 
(see bibl.~in \cite{LMKR99}). The shape of the continuum is determined by the 
distribution of population density among vibro-rotational levels of the upper 
$a^{3}\Sigma_{g}^{+}$ electronic state \cite{LP85}. It is not directly 
connected with any translational temperature but is determined by a dynamic 
balance between excitation and deactivation of the $a^{3}\Sigma_{g}^{+},v,J$ 
levels.

In low-pressure gas discharges, the spectral intensity distribution of the 
continuum may be calculated not only in relative \cite{LP85}, but even in 
absolute scale \cite{LP88} being in rather good accordance with experimental 
data. In the H$_{2}$+Ar mixtures the Ar$^*$(4s) $\to$ 
H$_{2}$($a^{3}\Sigma_{g}^{+}$) excitation transfer from long living 
(metastable and resonant) levels of Ar \cite{LM98} may play important role as 
well as the formation of excited ArH$^*$ excimer molecules and $A^{2}\Sigma$ 
$\to$ $X^{2}\Sigma$ spontaneous emission due to bound-repulsive transitions 
\cite{LMKR99,LFSM85}. The ArH continuum overlaps that of H$_{2}$ being located 
in the same wavelength range. The energy of He$^*$ and Ne$^*$ metastables is 
too high to participate in an excitation transfer leading to populating of the 
$a^{3}\Sigma_{g}^{+}$ state. However, Kr$^*$ and Xe$^*$ can do the job in 
three-body collisions with two 1S hydrogen atoms.

\begin{figure}
 \epsfxsize=8.2cm \epsfbox{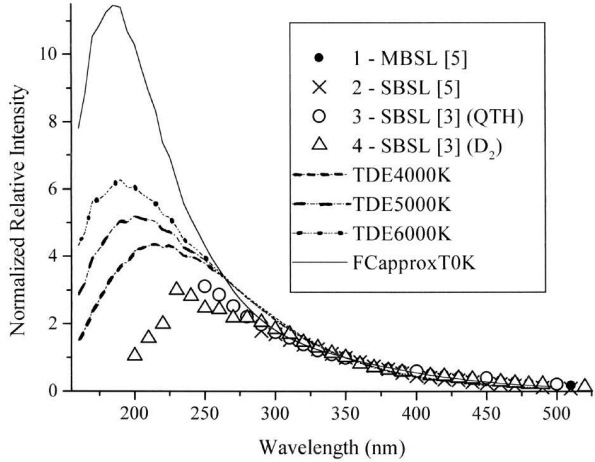}
 \caption{The comparison of recalculated experimental SL continuum spectra 
\protect\cite{HPB92,MRMMcNS95} with those calculated for the TDE conditions 
with $T$~=~4000,~5000 and 6000 K (dash-dot lines). Solid line corresponds to 
the case of direct electron impact excitation and spontaneous decay of 
$a^{3}\Sigma_{g}^{+},v$ levels of H$_{2}$ \protect\cite{LMKR99}.}
 \label{fig:4}
\end{figure}

Correct calculation of the continuum shape in plasma with temperature higher 
than 1000~K is impossible just because the transition probabilities are 
available only for rotation-less molecule \cite{LMKR99}. On the other hand, it 
needs development of certain model of microscopic excitation-deactivation 
processes and certain values of plasma parameters determined by macroscopic 
dynamics of collapsing bubble. Two main de-populating processes are for sure: 
spontaneous emission and radiationless collisional quenching \cite{BGP81}. But 
there is a great variety of competitive volume and surface processes 
responsible for the generation of H$_{2}^*$ ($a^{3}\Sigma_{g}^{+}$) excited 
molecules (electron impact, electron-ion and ion-ion recombination, 
associative three-body collisions, photo and/or collision-induced 
fragmentation of water, etc.). Nevertheless, very rough estimations can be 
made by neglecting the rotational structure of $a^{3}\Sigma_{g}^{+},v$ levels 
in two simple cases: 1) thermodynamic equilibrium (TDE) populations of 
$a^{3}\Sigma_{g}^{+},v$ levels relative to ground $X^{1}\Sigma_{g}^{+},v$=0 
vibronic state; 2) direct electron impact excitation and spontaneous decay of 
the levels \cite{LMKR99}. The results are shown in Fig.~\ref{fig:4}. One may 
see that calculated spectra are in accordance with experimental observations 
at least qualitatively. They have ``featureless structure'' with the intensity 
rising to UV cutoff $\lambda~\approx~250$~nm. In the range of observation, 
they may be fitted as ``blackbody spectrum'' with ``enormous temperatures''.

Experimental data are also not free from criticism:
   
1) Intensity calibration should take into account re-absorption in plasma and 
transparency of plasma-liquid boundary neglected in \cite{HPB92,MRMMcNS95}. 
   
2) Determination of the background level is not so simple, because light 
scattered inside a flask and a spectrometer overlaps dark signal of a 
detector. The huge difference between backgrounds of SBSL and MBSL curves is 
caused by use of two different optical systems in \cite{MRMMcNS95}. The 
emission of MBSL was focused on the entrance slit of the spectrometer by a 
lens. In the case of SBSL, no collection lens was used. The entrance slit was 
placed close to the side of the levitation cell being 2.25~cm from the bubble. 
It means that all light coming from within $2\pi$ solid angle (much bigger 
than the instrumental aperture and that for radiation directly coming from the 
bubble) was able to enter the spectrograph and partly be detected as a 
background.
   
3) The separation of the continuum intensity from the total signal of a 
detector is also a rather delicate and ambiguous deal. For example the 
peculiarities on the SBSL curve of Fig.~\ref{fig:2} may be interpreted as some 
additional emissions at $\lambda~\approx~310-360$~nm (OH bands) and 
$\lambda~\approx~400-500$~nm (NaH bands or H$_{2}$ continuum in the second 
order of the grating), or as absorptions at $\lambda < 320$~nm (OH) and 
$\lambda \approx~350-420$~nm (NaH).

4) Proper normalization of experimental curves is necessary for a comparison 
of their shapes. 
   
Experimental data of \cite{HPB92,MRMMcNS95} have been treated taking into 
account what is written above. Results of such recalculations are presented in 
Fig.~\ref{fig:4} \& \ref{fig:5}. One may see that the results of two 
independent SBSL experiments \cite{HPB92,MRMMcNS95} are in good agreement as 
well as intensity distributions obtained by MBSL \cite{MRMMcNS95} and SBSL 
\cite{HPB92,MRMMcNS95}.
   
Taking into account experimental errors and the uncertainties in the data 
processing we have to come to the following conclusion. After proper treating 
the experimental data show: 1) The continua emitted by SBSL and MBSL 
(Fig.~\ref{fig:5}) have identical spectral intensity distribution, therefore 
they may have the same nature; 2) Measured spectral intensity distributions 
and those roughly calculated for the $a^{3}\Sigma_{g}^{+},v,J \to 
b^{3}\Sigma_{u}^{+}$ spontaneous emission of H$_{2}$ molecule 
(Fig.~\ref{fig:4}) are in semi-quantitative agreement good enough to propose 
H$_{2}^*$($a^{3}\Sigma_{g}^{+}$) molecules to be responsible for the continuum 
emission. ``Enormous temperatures'' of SBSL reported so far have no physical 
meaning being the result of incorrect fitting (un-proper treated experimental 
data were approximated by un-proper analytical expression -- Planck formula).

There is actually a great difference between MBSL and SBSL experiments even if 
they are carried out with the same chemical solutions: 1) The amplitude of 
sound wave in MBSL (10~atm) is about one order of magnitude bigger than that 
used in SBSL ($\approx$1.3~atm). Therefore in MBSL the action of ultrasound 
should be much more powerful and destructive for bubbles. The widely 
distributed opinion that SBSL is a stronger phenomenon is based only on the 
``observation'' of ``enormous temperatures'', not more. 2) MBSL experiments 
are made with 100\% air saturation of a solution, while SBSL experiments are 
performed with degassed water. 

Thus, the bubbles have qualitatively different gas contents in those two types 
of SL experiments. MBSL bubbles are mainly air-filled with small amount of 
water vapor. Dissociation of N$_{2}$, O$_2$, H$_{2}$O during the collapse 
leads to formation of very aggressive species (like HN$_3$, HNO$_3$, 
N$_2$O$_2$, N$_2$O$_3$, etc.) which disappear by chemical reactions with water 
boundary. In the expansion phase a bubble (if it would be able to survive!) is 
again filled with air due to 100\% air saturation. New bubbles are definitely 
generated as air-filled. 

An absolutely other situation should occur in the case of SBSL when the action 
of acoustic waves is much more gentle and water is degassed. The SBSL bubble 
can accumulate not only Ar (1\% in air) \cite{LBDHJ97}, but molecular hydrogen 
as well. The hydrogen molecule in its ground state has almost the same 
electronic structure as that of He atom - its united atom analogue (two 1s 
electrons with anti-parallel spins). Thus, H$_{2}$ itself has low chemical 
activity in great contrast with hydrogen atom. The solubility of H$_{2}$ in 
water is much smaller than that of radicals made from N, O and H atoms. 
Therefore, a stable-oscillating bubble in SBSL mode actually consists of a 
H$_{2}$+Ar gas mixture with periodically changing amount of water vapor 
(increasing during the expansion and decreasing in the collapse). These 
additional H$_{2}$O molecules disappear in the collapse and serve as an engine 
(and fuel) for the transformation of the translational energy of collapsing 
liquid-gas boundary into the energy of light emission. This mechanism 
explains: 1) Why the light flash appears only at the first collapse but not at 
the second one in the series of damping oscillations in spite of almost the 
same compression \cite{P95}; 2) Why the average radius of a bubble generally 
increasing with a rise in acoustic amplitude suddenly shrinks when the onset 
of SL is reached \cite{P95}; 3) Great rise of SBSL with a decrease of water 
temperature \cite{HPB92}.
   
Clockwise regularity of SBSL flashes should not be so surprising and does not 
need unusual mechanisms because the experimental setup used in SBSL 
experiments is essentially a resonant system. The stability of this regularity 
means that after each collapse in the ``dead mode'' the bubble contents 
returns to the initial one - 98\% of (H$_{2}$+Ar) mixture and 2\% of H$_{2}$O.
   
The abnormally short duration of SBSL light flashes may be explained by 
extremely high rise of both the rate of excitation and the rate of collisional 
quenching of excited states. Thus the conditions suitable for spontaneous 
emission may be realized only in rather limited period of time. The situation 
is obviously different for different excited species. For some of them the 
favorable conditions could not be achieved at all (this explains also the 
existence of upper threshold of SBSL). The quenching may lead to a 
dissociation of molecules and to emission of vacuum UV radiation being out of 
the range of observation. The dim luminosity cloud surrounding a hot spot most 
probably is the fluorescence induced by L$_{\alpha}$ atomic line and/or Lyman 
and Werner bands of H$_{2}$. The estimation of the characteristic time of 
H$_{2}^*$($a^{3}\Sigma_{g}^{+}$) collisional quenching with cross sections 
from \cite{BGP81} gives $\approx$1~ps.
   
The positive influence of heavy inert gases Ar, Kr, Xe (in contrast to He, Ne 
\cite{DMcNS00}) may be connected with the excitation transfer from their 
metastables and with formation of excited hydrides like 
ArH$^*$($A^{2}\Sigma$). Absence of characteristic emissions of Na$^*$ and 
OH$^*$ in SBSL is caused by better evacuation and/or quenching of upper states 
in hydrogen-dominated contents of the bubble.

It is common practice in plasma spectroscopy that an investigator should find 
proper answers for three questions: 1) Who is the emitter of the emission? 2) 
What are main processes of excitation and deactivation of the upper state of 
the transition? 3) How the population density of the upper level(s) can be 
related to plasma parameters? Only when all three are answered the emission 
may be used for plasma diagnostics (spectroscopic determination of 
temperatures, particle densities etc.). From such point of view any 
speculations around abnormally high temperatures ``observed'' in SBSL and 
``the opportunity'' to make one more ``cold fusion'' are meaningless.

\acknowledgements 

My gratitude to M.~K\"aning for useful advices and help.

\clearpage

\begin{figure}
 \epsfxsize=8.2cm \epsfbox{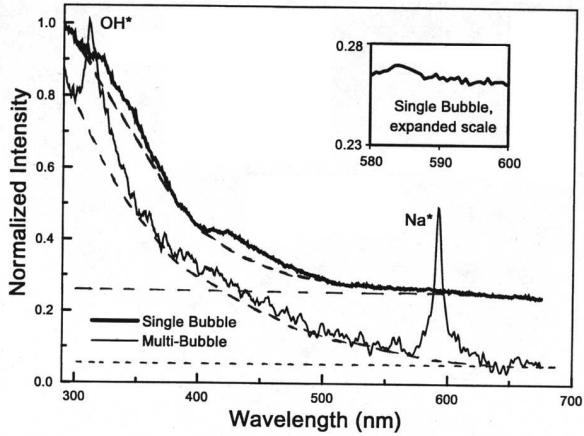}
 \caption{Spectra of SBSL and MBSL obtained in \protect\cite{MRMMcNS95} (solid 
lines). Dashed lines were used in present work as the continuum intensities 
with backgrounds shown by horizontal lines.}
 \label{fig:2}
\end{figure}

\begin{figure}
 \epsfxsize=8.2cm \epsfbox{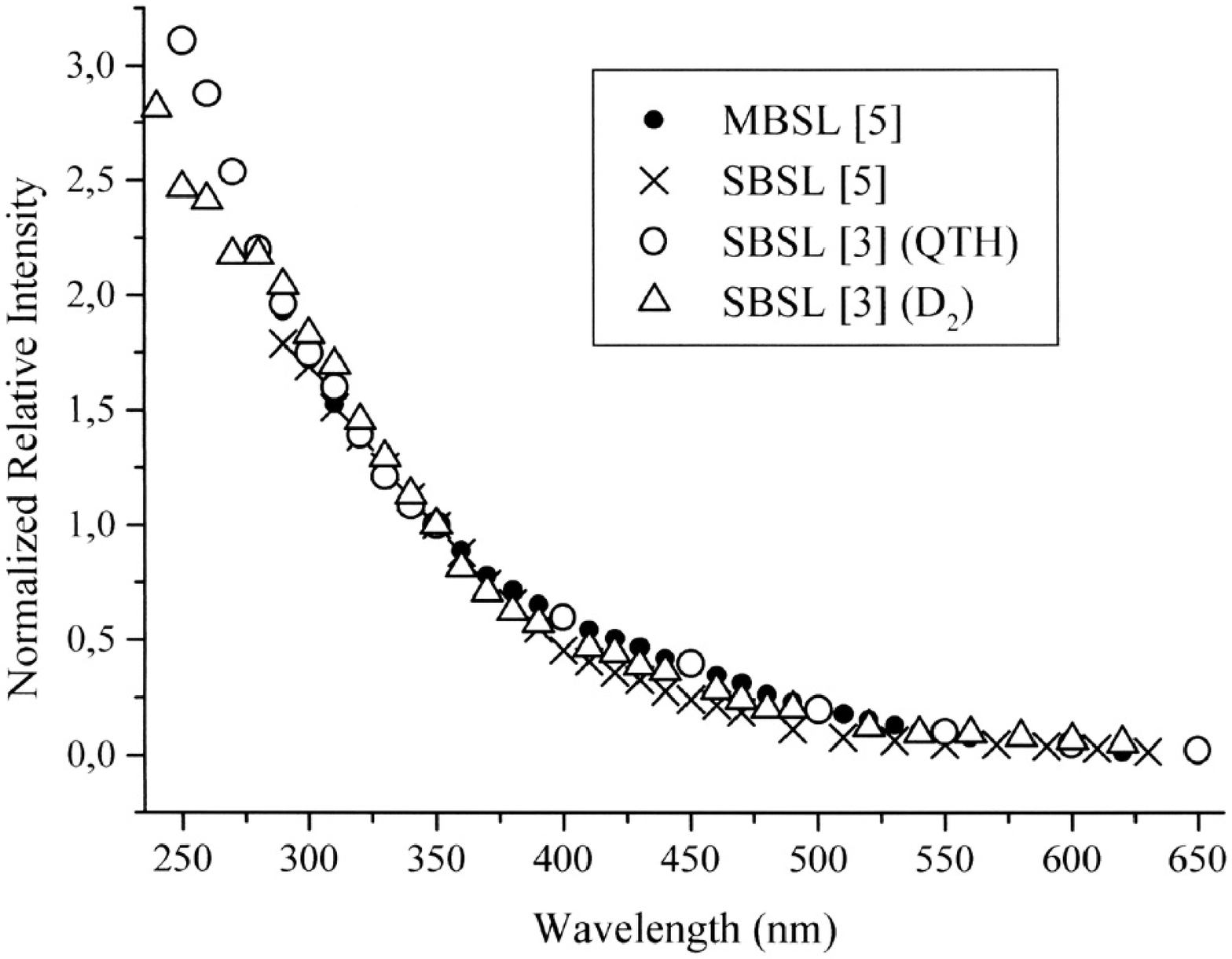}
 \caption{Relative continuum intensity obtained from SBSL and MBSL spectra 
\protect\cite{HPB92,MRMMcNS95} (Fig.~\ref{fig:1}, \ref{fig:2}) after the 
subtraction of the background and renormalization for unity at 
$\lambda$~=~350~nm.}
 \label{fig:5}
\end{figure}


\begin{thebibliography}{10}

\bibitem{Setal99}
S. Suslick {\it et~al.}, Phil.~Trans.~R.~Soc.~Lond.~A {\bf 357},  335  (1999).

\bibitem{McNDS99}
W.~B. {McNamara III}, J.~T. Didenko, and K.~S. Suslick, Nature {\bf 401},  772
  (1999).

\bibitem{HPB92}
R. Hiller, S.~J. Putterman, and B.~P. Barber, Phys.~Rev.~Lett. {\bf 69},  1182
  (1992).

\bibitem{GC90}
D.~F. Gaitan and L.~A. Crum,  in {\em Frontiers in Nonlinear Acoustics}
  (Elsevier, N.~Y., 1990), pp.\ 459--463.

\bibitem{P95}
S.~J. Putterman, Scientific American  33  (Feb.~1995).

\bibitem{C94}
L.~A. Crum, J.~Acoust.~Soc.~Am. {\bf 95},  559  (1994).

\bibitem{BHLPW97}
B.~P. Barber {\it et~al.}, Phys.~Lett. {\bf 281},  66  (1997).

\bibitem{MRMMcNS95}
T.~J. Matula {\it et~al.}, Phys.~Rev.~Lett. {\bf 75},  2602  (1995).

\bibitem{Y99}
K. Yasui, Phys.~Rev.~E {\bf 60},  1754  (1999).

\bibitem{LMKR99}
B.~P. Lavrov, A.~S. Melnikov, M. K\"aning, and J. R\"opcke, Phys.~Rev.~E {\bf
  59},  3526  (1999).

\bibitem{LP85}
B.~P. Lavrov and V.~P. Prosikhin, Opt.~Spectrosc. {\bf 58},  317  (1985).

\bibitem{LP88}
B.~P. Lavrov and V.~P. Prosikhin, Opt.~Spectrosc. {\bf 64},  298  (1988).

\bibitem{LM98}
B.~P. Lavrov and A.~S. Melnikov, Opt.~Spectrosc. {\bf 85},  666  (1998).

\bibitem{LFSM85}
C.~R. Lishawa, J.~W. Feldstein, T.~N. Stewart, and E.~E. Muschlitz,
  J.~Chem.~Phys. {\bf 83},  133  (1985).

\bibitem{BGP81}
J. Bretagne, J. Godart, and V. Puech, J.~Phys.~B: At.~Mol.~Phys. {\bf 14},  761
   (1981).

\bibitem{LBDHJ97}
D. Lohse {\it et~al.}, Phys.~Rev.~Lett. {\bf 78},  1359  (1997).

\bibitem{DMcNS00}
Y.~T. Didenko, W.~B. {McNamara~III}, and K.~S. Suslick, Phys.~Rev.~Lett. {\bf
  84},  777  (2000).

\end{thebibliography}
\end{document}